# Capillary flow of a suspension in the presence of discontinuous shear thickening

G. Bossis[1], Y. Grasselli[1,2] O. Volkova[1]

[1]*Université University Côte d'Azur, CNRS UMR 7010, Institute of Physics of Nice, Parc Valrose 06108 Nice, France*

[2]*Université University Côte d'Azur SKEMA Business School – 60 rue Dostoievski – CS30085 – 06902 Sophia Antipolis, France*

## Abstract

The rheology of suspensions showing discontinuous shear thickening (DST) is well documented in conventional rheometer with rotating tools, but their study in capillary flow is still lacking. We present results obtained in a homemade capillary rheometer working in an imposed pressure regime. We show that the shape of the experimental curve giving the volume flow rate versus the wall stress in a capillary can be qualitatively reproduced from the curve $\dot{\gamma}(\tau)$ obtained in rotational geometry at imposed stress but instead of a sharp decrease of the volume flow rate observed at a critical stress, this transposition predicts a progressive decrease in flow rate. The Wyart-Cates theory is used to reproduce the stress-shear rate curve obtained in rotational geometry and then applied to predict the volume flow rate at imposed pressure. The theoretical curve predicts a total stop of the flow at high stress, whereas experimentally it remains constant. We propose a modification of the theory which, by taking into account the relaxation of the frictional contacts in the absence of shear rate, well predicts the high stress behavior. We also hypothesized that the DST transition propagates immediately inside the capillary, once the wall shear stress has reached its critical value:$\tau_R=\tau_c$, even if the internal shear stress $\tau(r<R)$ is below the critical one. In this way the whole experimental curve can be well reproduced by the modified W-C model.

## Introduction

The rheology of concentrated suspensions is quite complex since it can exhibit a broad diversity of behavior like a yield stress, shear thinning or shear thickening or both and even sudden jumps of viscosity. These different behaviors are related to structural changes when the applied stress is increased. Depending on the balance between hydrodynamic shear forces imposed by the applied stress and local interparticle forces (lubrication, Van der Waals, Debye-Huckel, hydration, entropic etc..) individual particles can gather into different types of aggregates whose shape and life time will depend both on the applied shear rate and on the local interactions between the particles. A common feature is the fact that the suspending fluid imprisoned inside these aggregates move as if it was a solid part of the aggregates, then increasing the effective solid volume fraction and so the viscosity of the suspension. Such an approach can describe several rheological behaviors of concentrated suspensions (Quemada and Berli 2002, Bossis and Brady,1989). In monodisperse suspensions of colloidal particles a sudden jump of viscosity during a ramp of stress was attributed to the transition from a low viscosity configuration made of a  stacking of sheets of particles sliding over each other to a disordered one (Hoffman 1972). This transition is specific to monodisperse particles which can arrange in layers of hexagonally packed particles, thus increasing the average distance between sheared layers and lowering the viscosity compared to a disordered state.

Nevertheless, the presence of a sudden jump of viscosity was also observed in suspensions of polydisperse suspensions of Latex particles (Laun et al. 1991) and it was proved, by neutron scattering (Laun et al. 1992; Bender and Wagner 1996) that there was no layered pattern before the transition. Also suspensions of particles of irregular shape like corn-starch (Fall et al. 2008) or acicular calcium carbonate (Egres and Wagner 2005) or gypsum (Neuville et al. 2012) show this jump of viscosity whereas their irregular shape prevents the formation of ordered sliding layers. This discontinuous shear thickening (DST) was studied mostly in conventional rotational rheometry coupled with different techniques like dichroism (d'Haene et al. 1993), magnetic resonance imaging MRI (Fall et al. 2010), shearing of photo-elastic beads (Bi et al. 2011) neutron scattering (Laun et al. 1992; Bender and Wagner 1996) to obtain some information on the evolution of the structural properties during this transition. Both numerical simulations (Seto et al. 2013; Mari et al. 2014; Johnson et al. 2017; Singh et al. 2018; Guy et al. 2020) and experiments have shown  that the onset of the transition was ruled by the competition between the shear forces and the repulsive forces which prevent the surfaces to come in contact and to experiment friction forces. For instance, by varying the pH in suspensions of silica or alumina at a constant salt concentration, Franks et al. (Franks et al. 2000) have shown that an increase of the magnitude of the repulsive force was correlated with an increase of the shear stress needed to obtain the transition. More recently (Clavaud et al. 2017), using a suspension of silica spheres in a rotating drum to measure the friction coefficient, it was shown that the DST could only be observe if the suspension was initially in a frictionless state. There is now a large consensus that, in order to experimentally observe a DST transition, the particles should initially be separated by a repulsive force and that this transition will occur when the shear forces dominate the repulsive ones allowing a frictional contact between the surfaces of the particles. It is the formation of a percolated network of particles in frictional contacts, as the ones observed in numerical simulations (Mari et al. 2014; Gameiro et al. 2020), able to support the stress through elastoplastic contacts, which explains the jump of viscosity (see for example reviews (Brown and Jaeger 2014; Denn et al. 2018; Morris 2020)). The correlation between an increase of the normal force between particles and the one of the interparticle friction has been confirmed by AFM (Comtet et al. 2017; Hsu et al. 2018; Madraki et al. 2020).

A model which explains qualitatively this rheological behavior was proposed by Wyart and Cates (W-C) (Wyart and Cates 2014) and is based on the idea that the volume fraction at which the viscosity diverges, depends on the imposed stress. In the W-C model, this process was characterized by a jamming fraction $\Phi_J(\tau)$ lower than the usual jamming fraction of a perfect hard sphere suspension at random close packing (RCP), $\Phi_0$, but higher than a volume fraction $\Phi_J^\mu$ able to mechanically support a stress without flowing: $\Phi_J^\mu < \Phi_J(\tau) < \Phi_0$. This last limit reached for $\tau \to \infty$ is not well defined since it depends on macroscopic properties like the stiffness of the boundaries or the anisotropy of the structure and of local properties like the particle friction coefficient, $\mu$. At high friction ($\mu > 1$), a possible value for monodisperse spheres could be the loose random packing fraction $\Phi_J^\mu \sim \Phi_{RLP} \sim 0.55$-0.56 (Jerkins et al. 2008; Singh et al. 2018). In this model, for volume fraction between $\Phi_J^\mu$ and $\Phi_0$, it exists a domain of stress above which the suspension cannot flow, called the shear jammed domain. On the other hand, for $\Phi_c < \Phi < \Phi_J^\mu$ where $\Phi_c$ is the volume fraction under which there is no DST transition, the DST transition takes place with a S-shape of the shear stress versus shear rate curve, followed at high enough stresses by a Newtonian regime when $\Phi_J(\tau)$ becomes constant (Bi et al. 2011; Singh et al. 2018). These predictions are quite well verified by numerical simulations, but large disagreements persist when the parameters of the W-C model obtained by numerical simulations are transposed to represent the experimental data (Lee et al. 2020) or with polydisperse suspensions (Guy et al. 2020). An improved version of the model where the breakdown of the percolated force chains are allowed in the shear jammed zone was recently proposed (Baumgarten and Kamrin 2019). Also, attempts to introduce adhesive contacts, responsible for the presence of a yield stress, with a similar approach as in the W-C model, will help to generalize this model to a larger class of suspensions (Singh et al. 2019; Richards et al. 2020). The aim of this work is to bring some experimental data which can help to progress in the prediction of the rheological behavior of these concentrated suspensions, in particular with new experiments obtained with capillary flows. Usually, experimental data on DST phenomena are obtained in conventional rotational rheometry and very few papers present results obtained in capillary rheometry. A comparison between a cone-plate geometry and a capillary at imposed pressure on latex particles of diameter 0.3-0.4μm, has shown an apparent critical shear rate about two times larger in capillary geometry (Laun et al. 1991) which was attributed to wall slip. Pressure gradient in a square microchannel with confocal microscopy was used to study the velocity and concentration profiles of PMMA spheres of diameter 2.6 μm at $\Phi=0.63$; they found unexpected flow profiles, compared to yield stress fluids which were explained by taking into account the stress fluctuations (Isa et al. 2007). In situations of high confinements (R/a<30) a regime of oscillations of the flow rate was observed and explained by a local change of volume fraction associated with the permeation of the solvent through jammed domains (Isa et al. 2009). Recently, we have studied the flow of a concentrated magnetorheological suspension in a capillary at imposed flow rates (Bossis et al. 2020) showing that we recover the jamming transition but at a higher critical shear rate compared to conventional rheometry, both in the absence and presence of a magnetic field.

In the first section we shall present new experimental results for the flow of a suspension of magnetic particles in the regime of imposed pressure. The stress-shear rate curve obtained in conventional rheometry will be used to predict the volume flow rate at imposed pressure. The similarity between the predicted curve and the experimental one, despite the difference in flow geometry will be outlined but the reduction of the volume flow rate when the stress is increased is much more gradual than in experiments. In the second part we apply the W-C model to fit the experimental curve obtained in rotational rheometry and then we use it to predict the volume flow rate in the capillary configuration. As in rotational rheometry, the model predicts a total stop of the volume flow rate as the stress is increased contrary to what is observed experimentally. In the last section we propose a modification of the W-C model to explain the fact that the flow is not totally blocked above the transition, and we also show that it is necessary to consider a non-local rheology above the transition if we want to explain the abrupt transition observed in the capillary at imposed pressure.

## Materials and methods

We are using suspensions of carbonyl iron particles, grade HQ, from BASF, supplied by Imhoff & Stahl Gmbh; their density is $\rho_p$=7.8g/cm$^3$ and their average diameter d=2a=0.6μm. They are suspended in a mixture of 85% ethylene-glycol and 15% water which is used to minimize the evaporation at room temperature. The viscosity of the suspending fluid is $\eta_0$=0.011Pa.s. The additive used to prevent the aggregation between the particles is a plasticizer used in cement industry called Optima100 and sold by the company Chryso. The Reynold number based on the size of the particles is $Re_p = \rho\dot\gamma d^2/\eta$. For a typical shear rate of 10 s$^{-1}$ and the minimum viscosity of the suspension we have used: $\eta$=1Pa.s, the particle Reynold number is of order 10$^{-8}$, so inertial effects at the level of the particle are totally negligible in these suspensions. The sedimentation velocity of a particle $v = 2(\rho_p - \rho_f)ga^2/9\eta$ is of order 0.01μm/s so we can also neglect sedimentation effects in the capillary. Practically, all the commercial capillary rheometers impose the volumetric flow rate through the motion of a piston. In this case the DST transition results in a jump of pressure at a critical shear rate (Bossis et al. 2020) but more information on the physics of the transition can be obtained by driving the pressure since in this case, it is possible to observe the decrease of the shear rate and its subsequent behavior. For these experiments, we have used a homemade capillary rheometer described in Fig. (1).

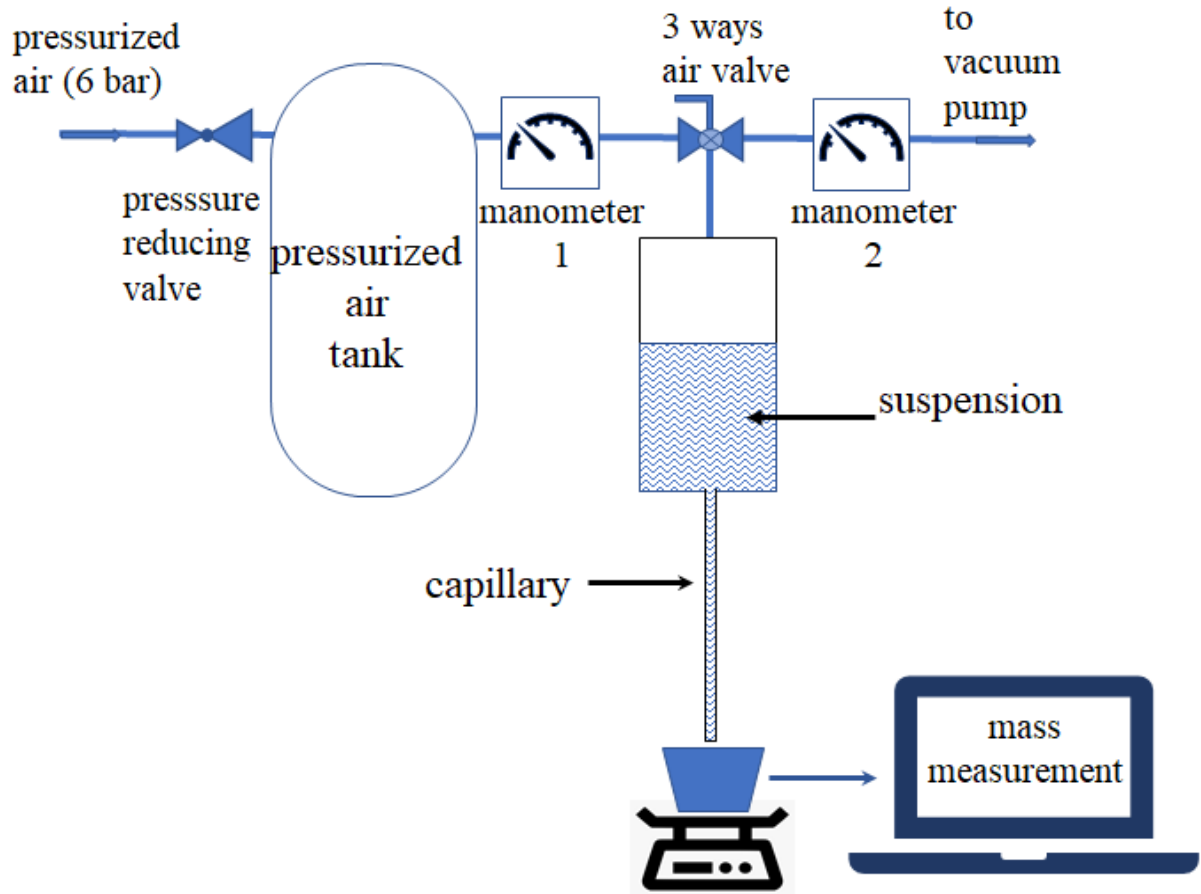

**Fig.1** Sketch of the equipment for capillary flow at imposed pressure.

It is composed of a tank with a pressure regulating valve (MDG-3 from Seflid) and a piezoelectric manometer (LEO 2 from Serv'instrumentation; resolution $10^{-3}$ bar, maximum pressure 3 bars) connected to a 3-way ball valve (SS-43GXS4 from Swagelok). For the gas, we use dry compressed air. The vertical output of the 3-way valve is connected to the liquid tank in plexiglass, itself connected to the capillary through an interchangeable flange whose output is a part of the capillary. This part can be prolongated through a tube fitting union (for instance SS-400-6 for a tube of ¼ in external diameter) to easily change the length of the capillary. The third way of the valve is connected to a water vacuum pump through a 2-way valve and a vacuum manometer. The suspension is first pumped from a container placed on a balance (STX from Ohaus) into the liquid tank at a known negative pressure and the change of weight recorded each second. Then the path of the 3-way valve is turned towards the pressurized gas tank, which is regulated at a given pressure, for instance during 60s, then the pressure is changed to another value and so on until emptying of the liquid tank; the change of weight is recorded during this time for typically ten different pressures. The procedure can be repeated several times until enough different pressure points have been recorded. The internal diameter of the tank is 2.5cm and its height 10cm. The radius of the capillary was R=1.5mm. The imposed pressure was corrected to consider the change of height of the column of fluid in the tank which was deduced from the change of mass. The mass flow rate Q(P) at a given pressure P, was obtained from the fit by a straight line of the mass versus time. The integration time of a mass measurement  was 1s, still we did not observe a decrease of flow rate at constant pressure which could have been attributed to clogging effects at the entrance of the capillary as predicted for small values of R/a (Koivisto and Durian 2017) since we have R/a~5000. The wall shear stress is related to the pressure by: $\tau_R = P.\frac{R}{2L}$ where L is the length of the capillary. In these experiments we have used two different lengths L= 36 cm and L= 18cm of a stainless-steel tube for the capillary and there was no difference between the data recorded for these two lengths.

On the other hand, the measurement of the viscosity was made with an imposed stress rheometer MCR 502 from Anton Paar. We used a cylindrical geometry with a small gap to minimize the change of shear rate inside the gap and to avoid particle migration.

## Experimental results.

The rheograms are represented in Fig. 2 for three volume fractions: Φ=0.6; Φ=0.64; Φ=0.65. For the two highest volume fractions, we have a strong signature of the DST transition with an abrupt decrease of the shear rate followed by strong oscillations around an average value which remains approximatively constant during the increase of stress.

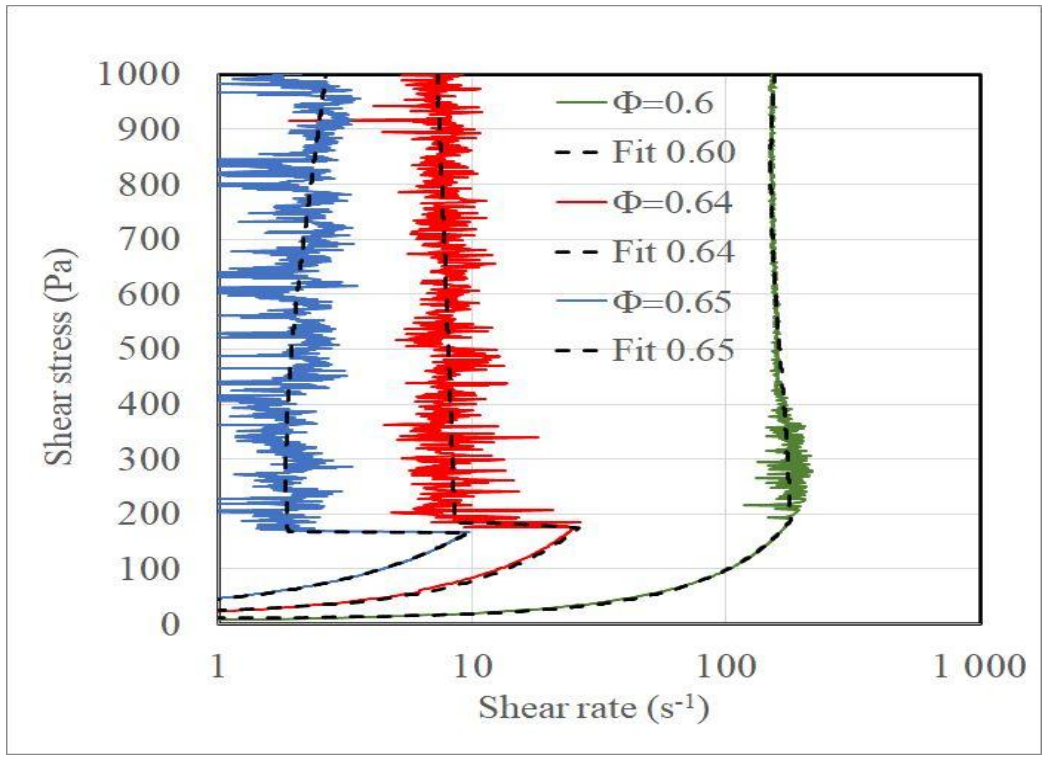

**Fig.2** Rheograms in cylindrical Couette cell for three volume fractions: Φ=0.6; Φ=0.64; Φ=0.65. The dashed black lines are a polynomial fit of the experimental curve.

The behavior is different at Φ=0.6 where the transition is still accompanied by oscillations of the shear rate but without a sudden decrease of its value. In a forthcoming paper focused on the rheology of MR suspensions at high volume fractions, we shall see that the volume fraction of 0.60 is approaching the one corresponding to the random loose packing of this suspension which is estimated to be $\Phi_{rlp}$=0.58 whereas the close packing is $\Phi_0$=0.681. The critical shear rate strongly decreases with the increase of volume fraction passing from 200s⁻¹ at $\Phi$=0.6 to 10 s$^{-1}$ at $\Phi$=0.65 whereas the critical stress decreases slightly. Above $\Phi_{rlp}$, the Wyart-Cates (W-C) model predicts that at high enough stress the suspension will be completely jammed and will stop. Actually, we rather observe, as many other authors (d'Haene et al. 1993; Laun 1994; Frith et al. 1996; Fagan and Zukoski 1997; Fall et al. 2015; Hermes et al. 2016), that above the transition, the shear rate remains approximately constant whatever the value of the stress. In practice at high stresses, it ends up by an expulsion of the suspension in plate-plate or cone-plate geometry or by foaming in cylindrical Couette geometry. We shall come back to the analysis of the W-C model in the context of the capillary flow in the next sections. The results obtained in capillaries for the wall shear stress as a function of the volume flow rates are presented in the three following figures (Figs 3-5). On these figures we have also plotted the predicted curve obtained from the fitted curve of the experimental rheogram $\dot{\gamma}(\tau)$ (cf Fig. (2)) through the equation:

$$Q(\tau_R) = \pi \int_0^R r^2\, \dot{\gamma}(r)\,dr = \frac{\pi R^3}{\tau_R^3}\int_0^{\tau_R} \tau^2\, \dot{\gamma}(\tau)\,d\tau \qquad (1)$$

This equation is always valid whatever the rheological law since it just comes from the definition of the volume flow rate as the integral from r =0 to r=R of the velocity field $v_z(r)$ and of the change of variable r=R.$\tau(r)/\tau_R$, where $\tau_R$ is the wall shear stress.

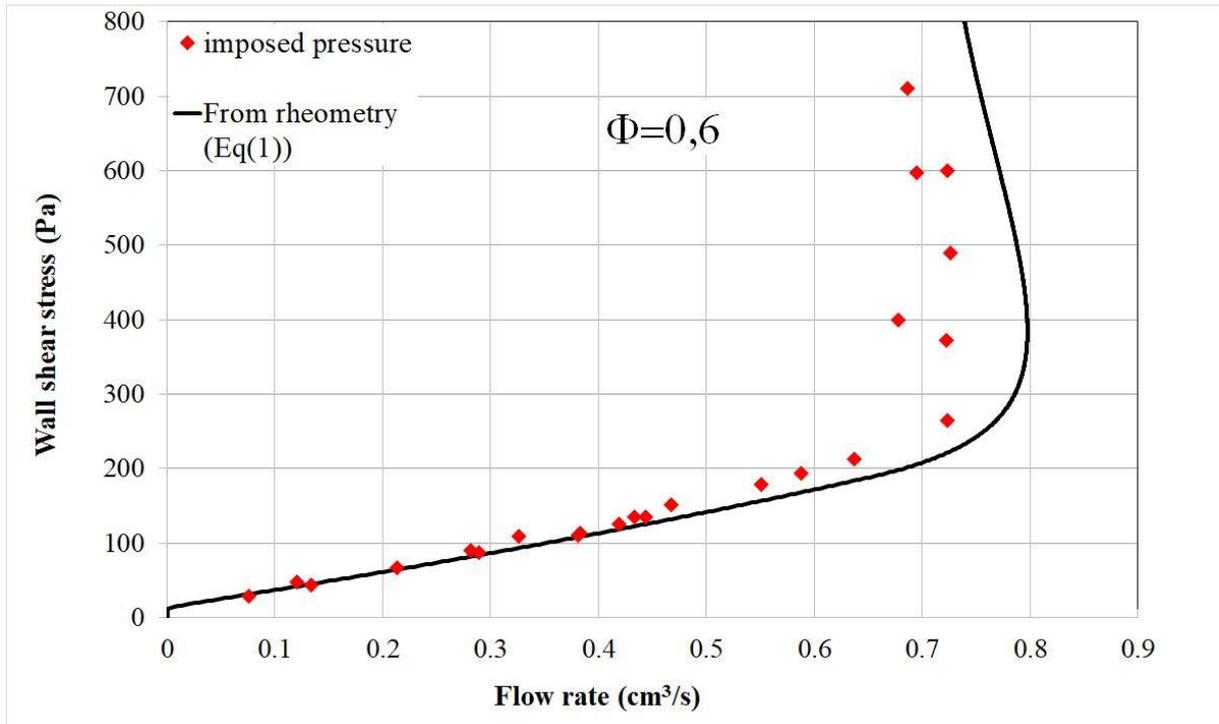

**Fig.3** Volume fraction $\Phi$=0.6. Red dots are the flow rates measured at different imposed pressures. The solid line is obtained from Eq. (1) with $\dot{\gamma}(\tau)$ given by the dashed curve of Fig. (2) for $\Phi$=0.6.

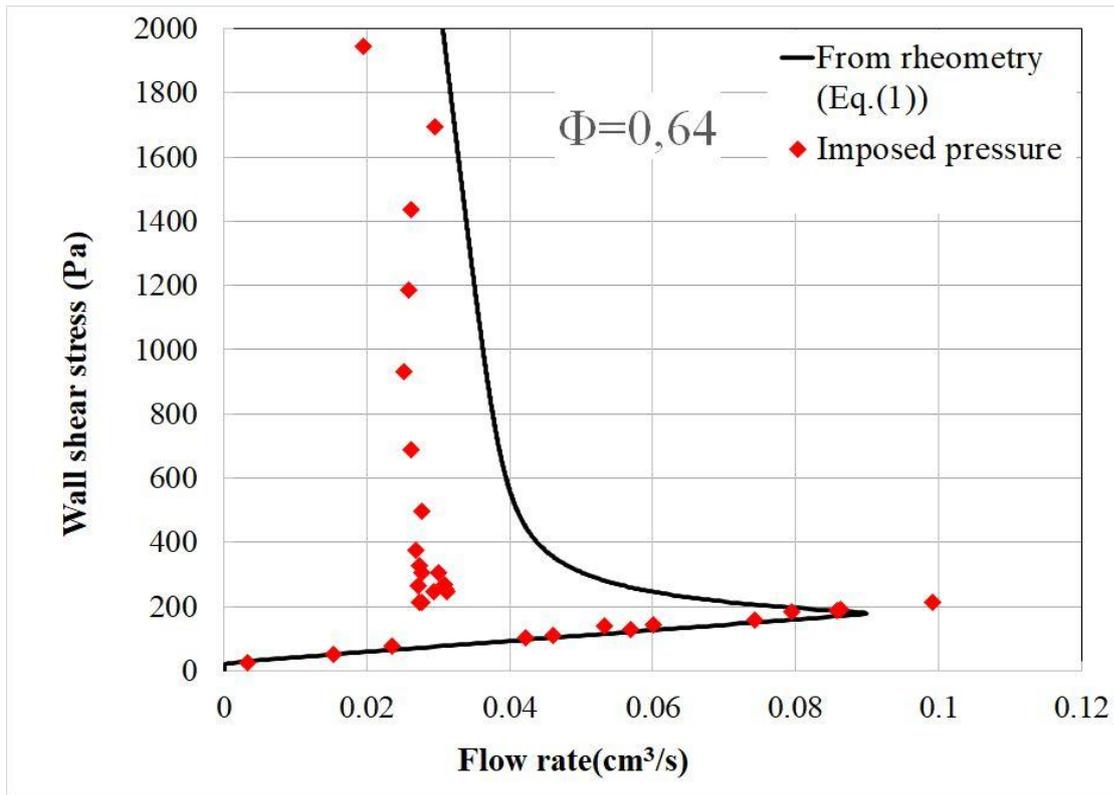

**Fig.4** Volume fraction $\Phi=0.64$. Red dots are the flow rates measured at different imposed pressures. The solid line is obtained from Eq. (1) with $\dot\gamma(\tau)$ given by the dashed curve of Fig. (2) for $\Phi=0.64$.

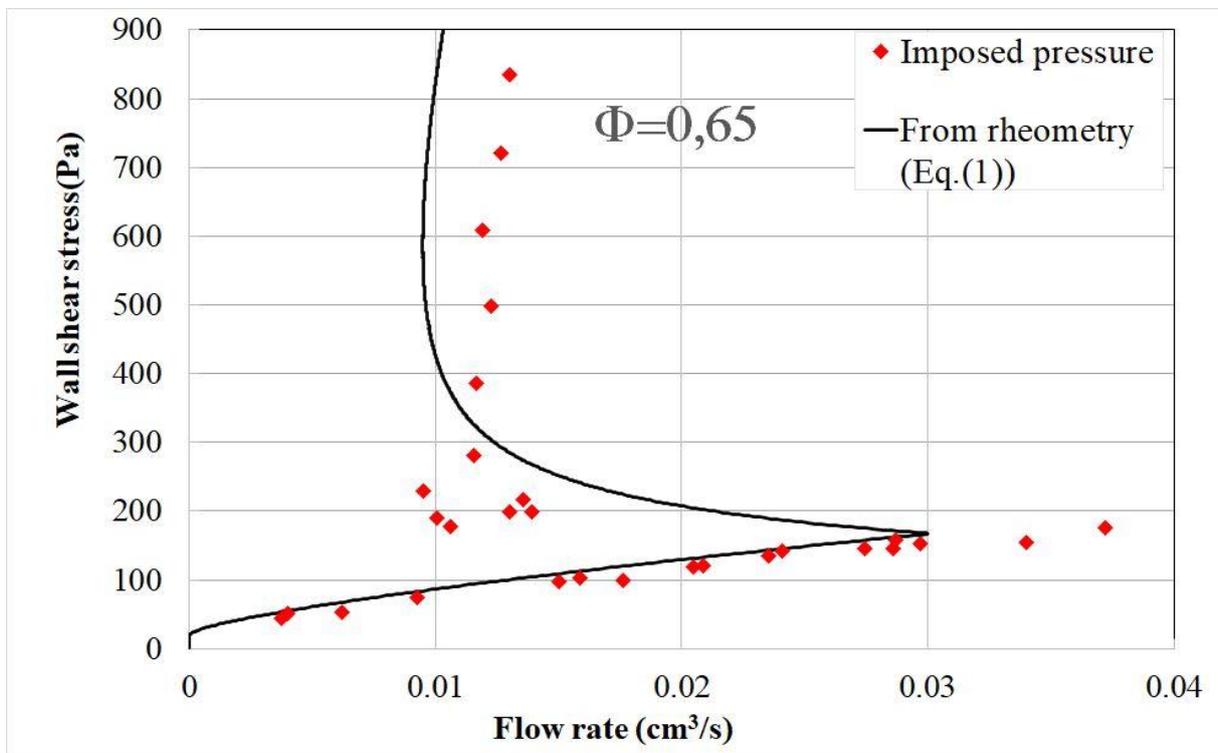

**Fig.5** Volume fraction $\Phi=0.65$. Red dots are the flow rates measured at different imposed pressures. The solid line is obtained from Eq. (1) with $\dot\gamma(\tau)$ given by the dashed curve of Fig. (2) for $\Phi=0.65$.

The previous figures show that , qualitatively, the predicted result from the use of rheometry through Eq.(1) is in agreement with the experimental points showing a decrease of the flow rate above a critical flow whose value is nevertheless smaller than the experimental one for $\Phi$=0.64 and $\Phi$=0.65. At $\Phi$=0.60, where the rheometry does not show a sudden decrease of the shear rate (Fig. (2)), the predicted curve has the same aspect as the experimental one with a rounded part at the transition. This is quite different for $\Phi$=0.64 and $\Phi$=0.65 (Figs 4 and 5) because the predicted curves do not reflect the sudden decrease of the flow rate

We have seen that both in rotational rheometry and in capillary flow we have a sharp transition but that, if we introduce the experimental curve $\dot{\gamma}$ ($\tau$) obtained from rheometry in Eq. (1), the predicted transition for the capillary flow is a soft one; (cf. Figs. (4)-(5)). Still Eq.(1) is general and applies whatever the rheological law. We will see what the reason is for this contradiction in the last section, but first let us see how the W-C model applies to a capillary flow.

## Application of the Wyart-Cates model

In the W-C model the volume fraction where the viscosity diverges, depends on the applied stress in the following way:

$$\Phi_J(\tau) = (\Phi_{\text{RLP}} - \Phi_0)f(\tau, \lambda, q) + \Phi_0 \qquad (2)$$

As explained in the introduction $\Phi_{\text{RLP}}$ can be approximated by the random loose packing which is 0.58 for our suspension and the frictionless packing is $\Phi_0$=0.681. The function $f$ represents the proportion of frictional contacts in the suspension and increases from 0 to 1 while increasing the stress. Different expressions have been proposed for the function $f$; we shall take the following one (Guy et al. 2020):

$$f(\tau^*, \lambda, q) = e^{-\left(\frac{\lambda}{\tau^*}\right)^q} \text{ with } \tau^* = \tau/\tau_c \qquad (3)$$

The parameters $\lambda$ and $q$ are taken to fit the experimental curve obtained in conventional rheometry and $\tau_c$ is the critical stress at the transition.

The viscosity and the shear rate dependence are given by the usual relation for concentrated suspensions, with the only difference that $\Phi_0$ has been replaced by $\Phi_J(\tau)$. The yield stress in Eq. (4) is the dynamic yield stress of a Bingham law which fits quite well the beginning of the stress-shear rate curve:

$$\dot{\gamma}(\tau) = \frac{\tau - \tau_y}{\eta_c(\Phi, \tau)} \text{ with } \eta_c(\Phi, \tau) = \alpha(1 - \frac{\Phi}{\Phi_J(\tau)})^{-2} \qquad (4)$$

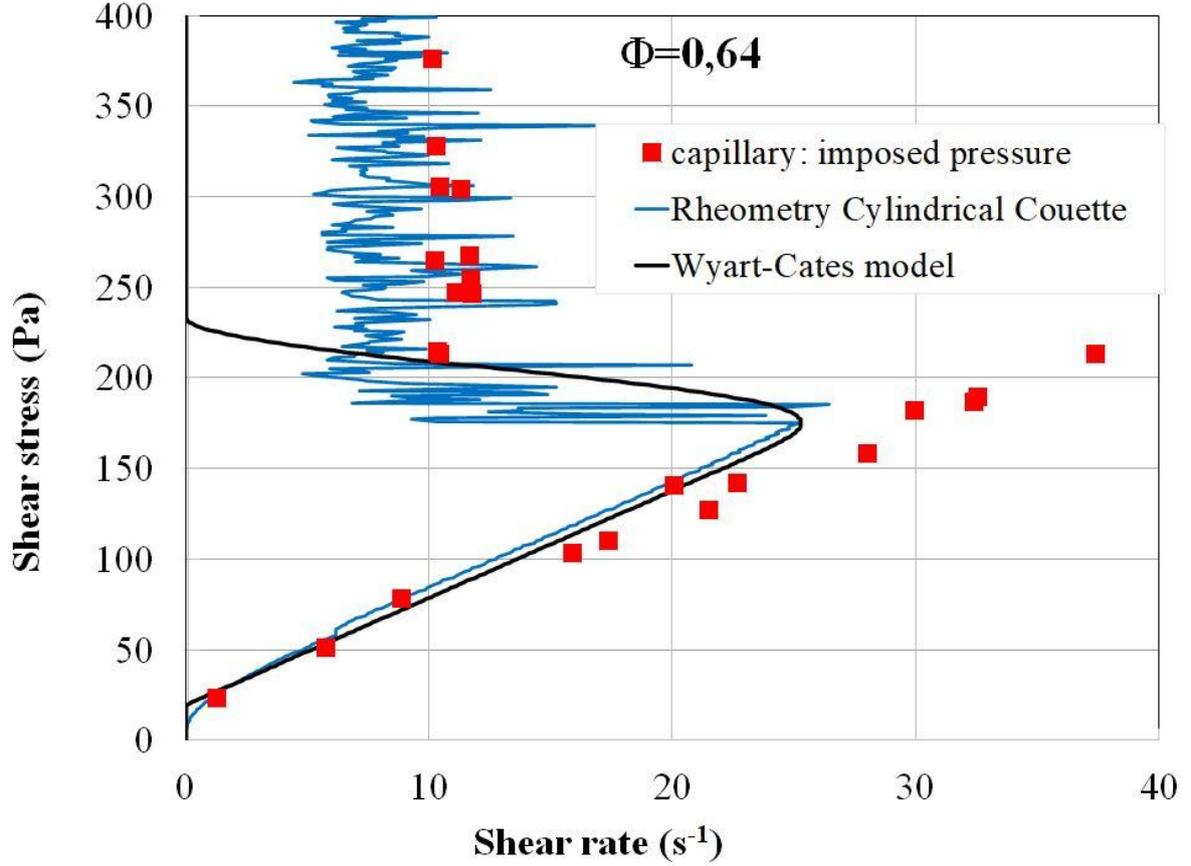

**Fig.6** Fit of the rheometric curve (blue solid line) by the Wyart-Cates model (solid black line). The shear rate for the results at imposed pressure (red squares) is deduced from the flow rate by: $\dot{\gamma} = 4Q/(\pi R^3)$

The values of $\alpha$ and $\tau_y$ are obtained from the fit by Eq. (4) of the first part of the curve for $\tau < \tau_c$ and $\lambda$ and q are obtained by the condition that the fitted curve should have a zero slope for $\frac{d\dot{\gamma}}{d\tau}$ at the experimental turning point $\tau_c, \dot{\gamma}_c$. The result is represented in Fig. (6) by the solid black line with the following values:

$\tau_y$= 19.4 Pa, $\alpha$=0.02, $\lambda$=1.29, q=6.08

We see that the W-C model predicts a zero shear rate when the jamming fraction $\phi_J(\tau)$ becomes equal to the actual value, $\Phi$, whereas as already pointed out, the experimental shear rate does not cancel and remains approximatively constant (solid blue line). It is also interesting to note that on this figure, the shear rate for the results at imposed pressure is the Newtonian shear rate $\dot{\gamma}_N = 4Q/(\pi R^3)$. This shear rate is not the real wall shear rate which is obtained by the derivative of Eq. (1):

$$\dot{\gamma}_R = \frac{1}{\pi R^3}\left[\tau_R \frac{dQ(\tau_R)}{d\tau_R} + 3Q(\tau_R)\right] \tag{5}$$

The derivative of the volume flow rate relatively to the stress is not well defined experimentally, especially during the transition; this is the reason why we have decided to compare in Figs. (3-5), the wall stress versus volume flow rate rather than versus the shear rate. The quite large difference between the critical shear rate in conventional rheometry and in capillary flow in Fig. (6) is partly due to the use of the Newtonian shear rate instead of the real one, and we can see in Fig. (4) that the agreement is better if we consider the volume flow rate instead of the Newtonian shear rate.

Since, the Eq. (4) represents the W-C model for the function $\dot{\gamma}(\tau)$ in rotational rheometry, we can use this analytical equation in Eq.(1) to compare the prediction of this model with the experiment in the case of a capillary flow. This is the solid blue line in Fig. (7).

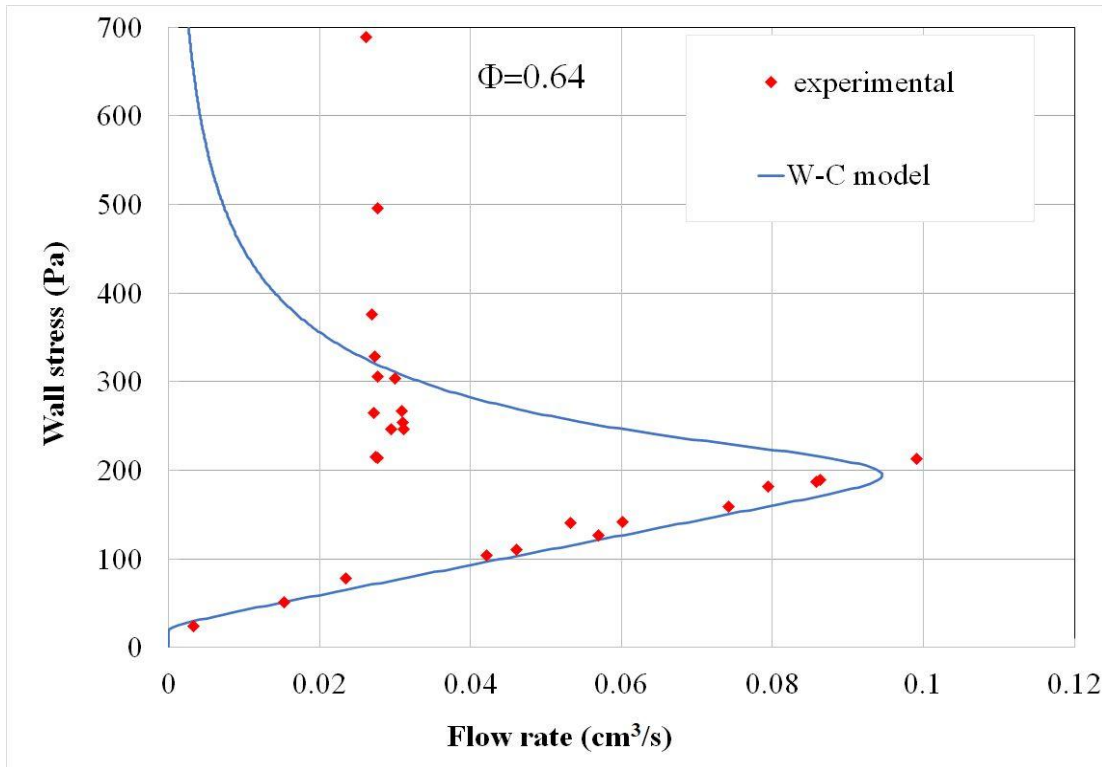

**Fig 7** Prediction of the stress versus flow rate in the frame of the W-C model. The experimental points for the capillary flow are the red dots at $\Phi=0.64$.

As this model predicts a complete jamming above a given stress ($\tau_m=233$Pa for $\Phi=0.64$), the predicted volume flow rate will also progressively stop when the stress rises. In this model, since the stress remains low at the center of the capillary, the central part will continue to flow even if a solid zone has begun to form on the wall where the stress is larger than the jamming stress $\tau_m$. This zone will progressively extend towards the center until it fills all the section of the capillary when the flow stops. There are two striking differences between the prediction of the W-C model and the experiment. The first concerns the soft transition instead of the abrupt one and was expected since it was already observed with the fitted experimental rheometric curve that the W-C theory try to represent. The second is that the volume flow rate decreases towards zero while experimentally it remains constant above the transition.

A possible reason for the soft transition could be related to shear induced migration which would gradually modify the volume fraction profile and could then be responsible for this behavior. Actually, shear induced migration above DST was observed in a Couette cell of large gap by magnetic resonance imaging with cornstarch particles (Fall et al. 2010) so we have to analyze its possible role on the stress-volume flow rate curve.

In the appendix we develop an analysis of shear induced migration in the presence of aggregates based on the work of Mills and Snabre (Mills and Snabre 1995). The predicted effect of migration on the wall stress versus flow rate curve is shown to increase the volume flow rate between 50% and 100% at the vicinity of the critical stress compared to the experimental values which on the other hand, are well predicted in the absence of migration (cf Appendix, Fig 11). The lack of migration could be surprising at first glance given that it was observed with cornflake particles by magnetic resonance imaging in a large gap Couette cell where the shear rate is also inhomogeneous. In fact, our particles are more than an order of magnitude smaller and the diffusion coefficient being proportional to the square of the radius of the particles we expect a much smaller migration as long as the particles are not aggregated. On the other hand, above the transition, in the presence of a percolated network of frictional contacts, the transverse motion of particles during the deformation and rearrangement of this network is probably a rarer event than when particles do not stick to each other. This analysis shows that shear induced migration can't explain why the use of the rheometric data obtained in rotational geometry, as well as the W-C model based on these data, fail to reproduce the sharp change of volume flow rate in capillary flow. In the last section we will propose an explanation of this disagreement, but first we need to investigate why the volume flow rate does not stop above the transition as predicted by the W-C model.

## Modification of the W-C model to get a constant flow rate above the transition

If the flow stops, we expect that the normal force will relax either through remaining Brownian motion or lubricated contacts still present in the percolated network or because the polymer molecules which were strained or removed from the surface by the local flow will come back to their initial position. In this case, the percolated structure will be momentaneously isotropized and destroyed, but the high applied stress will make the suspension to flow again and the cycle between arrested flow and flowing states will start again. This qualitative explanation is based on an equilibrium between forces which on one hand tend to destroy the percolated network of frictional contacts and on another hand to reinforce it. This approach is generally used to describe the evolution of a structural parameter like for instance, the sizes of the aggregates in the presence of shear and of attractive or Brownian forces and to predict the rheological law related to this structural parameter (Quemada and Berli 2002). In this kind of approach, the flowing state observed above the transition can be the result of an average of unstable flows which depends on the ratio between the measurement time and inertial time $t_m/t_I$. Another way to describe this phenomenon is to say that the structure at the jamming volume fraction $\Phi_J(f)$ (which is smaller than the RCP one, $\Phi_0$), can still deform and yield so, depending on the difference between the applied stress and the yield stress generated by the percolation of frictional contacts, the suspension can still flow through rupture and reformation of frictional bonds. In a recent paper (Baumgarten and Kamrin 2019), a similar reasoning was applied to the fraction, *f*, of frictional contacts (here defined by Eq. (3)). A constant shear rate in the limit of high stress was obtained thanks to the introduction of a "hardening function" $H(\tau)$ scaling as $\tau^{3/2}$. A simpler way to obtain a constant shear rate when the jamming fraction approaches the actual volume fraction is to introduce a divergence of the viscosity at a given non-zero shear rate. As already said, the fraction of frictional contact must tend to zero with time when the suspension no longer flows and that, whatever the value of the initial stress. Let us then write the fraction of frictional contact as:

$$f'(\tau^*, \dot{\gamma}) = f(\tau^*) * L(t_L.\dot{\gamma}) \qquad (6)$$

Here $f(\tau^*)$ is the function already defined by Eq. (63) and L(x) is a function which tends to zero when x tends to zero and which saturates to unity like for instance the Langevin function: L(x)=coth(x)-1/x. The parameter $t_L$ is related to the relaxation time of the stress during the blockage phase: the larger, the closer we approach a complete stop of the flow. We call $f_{jam}$ the value of $f$ for which we have $\Phi_J(f_{jam})=\Phi$ (here using Eq. (2) and (3) give $f_{jam}$=0.44 for $\Phi$=0.64). Since at high stresses $f(\tau^*) \to 1$, the equation $L(t_L.\dot{\gamma}) = f_{jam}$ will give the limiting shear rate, $\dot{\gamma}_l$, for which the viscosity $\eta(f')$ diverges; in our example for $\Phi$=0.64 and $t_L$=0.18, $\dot{\gamma}_l = 8.49\ s^{-1}$. Due to this divergence, the shear rate $\dot{\gamma} = (\tau - \tau_y)/\eta(\tau^*, \dot{\gamma})$ approaches quickly its asymptote. We have added here the presence of a dynamic yield stress which is present in our case and the viscosity is the one given by Eq. (4) with $f'$ instead of $f$. The resulting curve is shown in Fig. (8) with the relaxation time $t_L$=0.18s together with the original prediction of the W-C model. We see that this simple modification allows to well represent the experimental data.

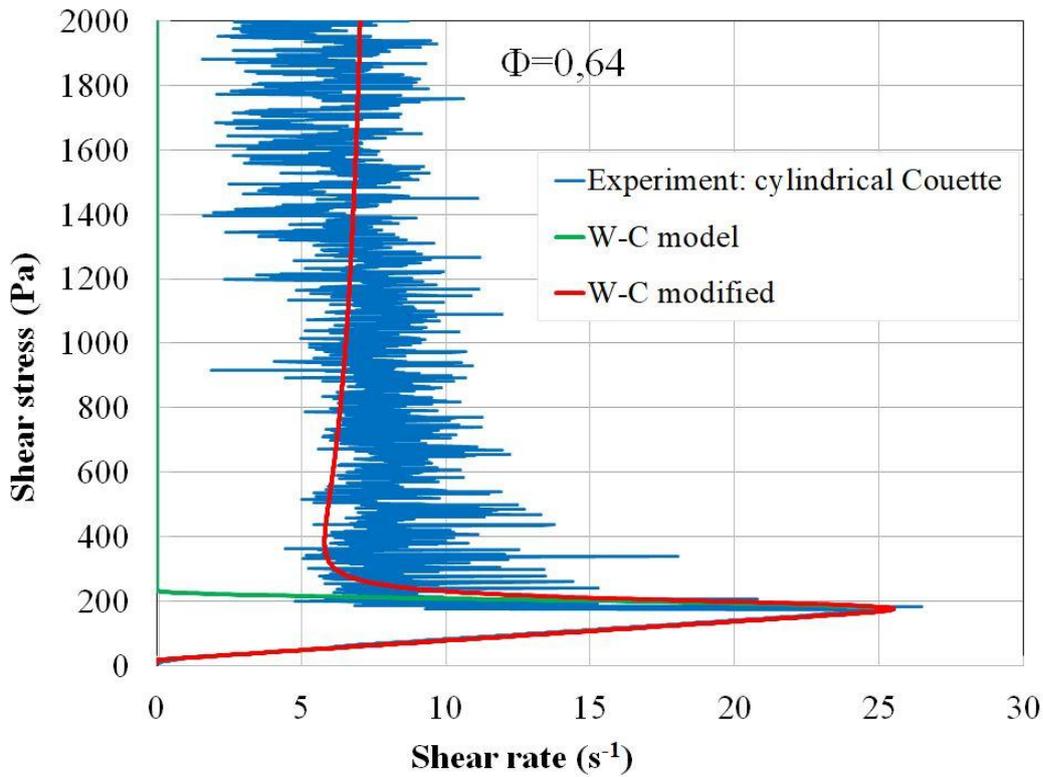

**Fig.8** Comparison of the prediction of the W-C model (green line) with the modified value of $f$: Eq. (6) and $t_L$=0.18s (red line). The solid blue line is the experimental result in Couette geometry at $\Phi$=0.64.

As it is not the scope of this paper we did not try to reproduce the oscillations of the shear rate which can be obtained by including the inertia of the rotating part of the rheometer in the equation of motion of the tool (Bossis et al. 2017, 2019; Richards et al. 2019). We do not see any fluctuations of the flow rate at imposed pressure although they could appear due to the inertia of the fluid itself (Nakanishi et al. 2012) but the averaging time of 1s for the mass measurement is too large to observe them if any, because the inertia time: $\tau_I = \rho R^2 / 2\eta$ ,which rules the relaxation of a shear rate fluctuation, is typically smaller than 1ms in our experimental conditions. It should also be noted that, for the analytical curve $\dot{\gamma}(\tau)$ derived from the rotational rheometry data, we have used an average value of the fluctuating shear rate, implicitly assuming that this average was the equilibrium curve in the absence of inertia and that it is the right value to introduce in Eq. (1) since, in the capillary flow, the inertia is negligible and its effect is not observed. Now that we have an equilibrium curve from the modified W-C model we could also use the theoretical curve $\dot{\gamma}(\tau)$ of fig. (8) instead of the fitted curve in Eq. (1) but the result would be very close from the one obtained with the fitted rheometric curve: there will be no sharp transition of the volume flow rate as observed experimentally. So, the modification of the W-C theory does not help to explain this discrepancy.

On the other hand, Eq. (1) is always valid, but we have done the hypothesis that the rheological law inside the capillary was the same as in the cylindrical shear cell, or in other words that the law $\dot{\gamma}$ ($\tau$) fitted from the rheometry applies locally inside the capillary with $\tau = (r/R)\tau_R$. This is not very realistic: as soon as the transition takes place close to the wall where the shear stress is maximum, the radial stress will provoke the propagation of the percolation towards the center of the cell. We can assume that above the transition, we suddenly move towards a regime where we have a uniform viscosity given by the modified W-C model. Said differently, the jamming volume fraction is the one at the wall $\Phi_J(\tau_R)$ and does not depend on the radial coordinate. Assuming a constant viscosity independent of the radial coordinate means that we have now in Eq. (1):

$$\dot{\gamma}(\tau) = \frac{\tau}{\eta(\tau_R)} \quad \Longrightarrow \quad Q(\tau_R) = \frac{\pi R^3}{4}\dot{\gamma}(\tau_R) \tag{7}$$

Above the transition, even if the percolated structure extends to the center of the capillary, it is likely that the contact forces between the particles will be lower at the center and consequently that the viscosity will also be lower; nevertheless, as it can be seen in Fig. (9), this approximation represented by Eq. (7) well reflects the experimental behavior.

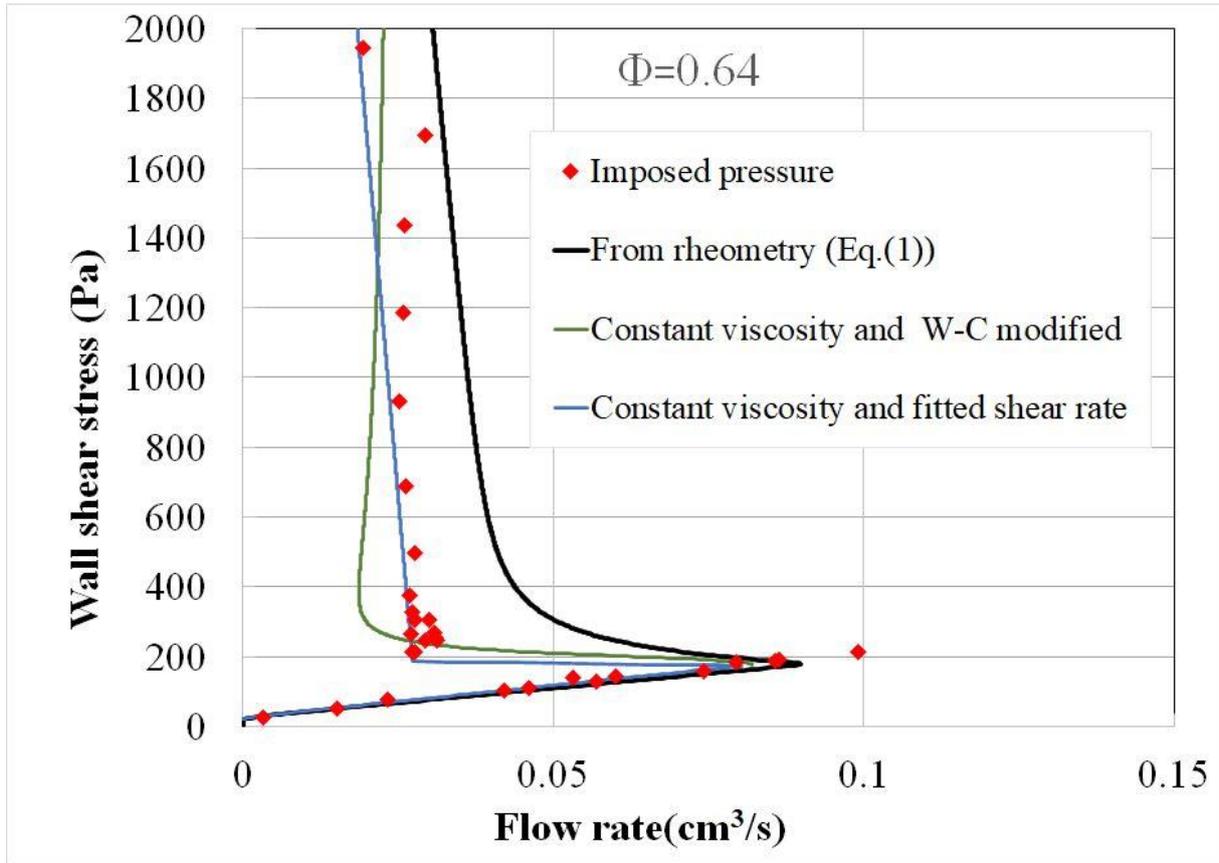

**Fig.9** Different models used to predict the volume flow rate at imposed pressure. Solid black line: Eq. (1) with fitted value of $\dot{\gamma}(\tau_R)$ (cf. Fig. (2)). Green solid line: Eq. (7) with $\dot{\gamma}(\tau_R)$ from the modified W-C model. Blue solid line: Eq. (7) with the fitted values of $\dot{\gamma}(\tau_R)$. Red dots: experimental points at $\Phi=0.64$.

The green curve is the result of Eq. (7) applied to the prediction of $\dot{\gamma}(\tau_R)$ by the W-C model modified with the help of Eq. (6) and the blue curve is the same Eq. (7) applied to $\dot{\gamma}(\tau_R)$ fitted from the rheometry cf. (Fig. (2)). Even if this approach is an "oversimplification", the important point to underline is that the comparison between the data obtained in the capillary at imposed pressure and those obtained in Couette cell geometry can't be explained above the transition without abandoning the hypothesis of a local rheology represented by the solid black line in Fig. (9). This point would be quite trivial for small value of R/a but here we have R/a ~5000 proving the long-range effect of the transmission of normal forces above the DST transition.

## Conclusion

We have presented new experiments based on capillary flow at imposed pressure of a suspension in the regime of discontinuous shear thickening. The rheological law $\dot{\gamma}(\tau)$ of the same suspension was measured in conventional rheometry using a cylindrical Couette cell. At high enough volume fraction, we well recover the DST transition in this capillary flow with a strong and abrupt decrease of the volume flow rate for a critical stress. The experimental curves $Q(\tau_R)$ of the volume flow rate as a function of the wall stress were compared with the ones predicted using the rheological law from Eq. (1). The agreement was qualitative but did not reflect the abrupt transition observed experimentally. A possible reason could have been the migration of particles towards the center, but an analysis of shear induced migration shows that it would give a volume flow rate much larger than the experimental one. The use of the Wyart-Cates model to predict the volume flow rate does not help and furthermore predicts a flow stop above the critical stress which is not observed experimentally. On the contrary, as in conventional rheometry for the shear rate, the volume flow rate remains approximatively constant when the stress is increased. It is possible to explain this behavior considering that, if the flow stops for a given stress, the absence of shear flow will allow to reactivate the interparticle repulsive forces and will lead to a deformation of the percolated structure and a new start of the flow. Allowing the fraction, *f*, of frictional contacts to cancel at low shear rate independently of the value of the applied stress (cf. Eq. (6)) allows us to well reproduce the zone of constant flow rate at high stress. The other discrepancy stands in the fact that, the use of the curve $\dot{\gamma}(\tau)$ obtained in a Couette cell to predict the one observed in the capillary shows a gradual decrease of the volume flow rate above the DST transition rather than the abrupt one observed experimentally. If we consider that, in the presence of a percolated network of contacts, the viscosity does not depend on the radial position  and is equal to the one on the wall: $\eta(\tau_R)$, (or in other words that the frictional network has extended over the entire section of the capillary), then it is possible to recover the volume flow rate obtained at imposed pressure from the rheometry in a Couette cell.

## Acknowledgment

The authors want to thank the CENTRE NATIONAL D'ETUDES SPATIALES (CNES, the French Space Agency) for having supported this research

## Appendix: Effect of Shear induced migration on the volume flow rate

Shear induced migration in a gradient of shear rate is related to the gradient of the number of collisions between particles since each collision induces a random transverse motion of order of magnitude *a*, relatively to the average velocity direction which induces a migration from the higher rate of collision domain (high $\dot{\gamma}$) to the lower one. This migration was first observed by NMR in concentric cylinders (Abbott et al. 1991; Graham et al. 1991; Chow et al. 1994) and in Poiseuille flow (Hampton et al. 1997). The shear induced diffusion coefficient is proportional to $\dot{\gamma}a^2$ with a coefficient which depends on the volume fraction and of the interparticle force but that remains usually smaller than unity for reasonable values of the range of roughness (Leighton and Acrivos 1987; Da Cunha and Hinch 1996; Zarraga and Leighton Jr 2001) or of the interparticle forces (Meunier and Bossis 2008). In our case with an average radius of 0.3μm,  the typical time for a migration of 1mm - the characteristic value of the radius of the capillary- would be $T_m \propto \frac{1}{\dot{\gamma}}\left(\frac{R}{a}\right)^2 \approx 10^5 s$ for a typical wall shear rate of 10s[-1]. The time needed for the transfer of the suspension from the input to the output of the capillary is $T_{tr} = \pi R^2 L/Q$ then, with a radius of 1.5mm and the maximum length used of 36cm and a minimum flow rate of 0.01cm[3]/s, we get $T_{tr}$=250s which is more than two orders of magnitude smaller than the migration time. Nevertheless, if below the transition we can neglect the migration, this not so obvious anymore above the transition since it is related to the formation of aggregates of particles in frictional contact which are supposed to percolate between the walls of the capillary. Instead of the radius of the particles which determines the transverse change of trajectory, it is rather a typical size of the aggregates, which means that now the migration time becomes just proportional to $\dot{\gamma}^{-1}$ which in turns, is now much smaller than the transfer time of the suspension. Of course, this analysis is only indicative because the collision between big aggregates will mainly result in their deformation rather than in a global change of trajectory. As already underlined, the size of the aggregates formed during the transition can play a major role in shear induced migration. We shall use the work of P. Mills and P. Snabre (Mills and Snabre 1995) who have introduced in a simple way a correlation length ,$\xi$, in order to estimate its influence on the migration. They obtain the following result for the volume fraction $\Phi(r)$ in a circular capillary:

$$\frac{\Phi(r)}{1-\Phi(r)/\Phi_0} = \left(\frac{r}{R}\right)^L \frac{\Phi_R}{1-\Phi_R/\Phi_0} \tag{8}$$

Where $\Phi_0$ is the closed-packed volume fraction,$\Phi_R=\Phi(R)$ and $L$ an exponent which is equal to 2 for 0<r<$\xi$ and 1 for $\xi$<r<R. The correlation length $\xi$ is the one over which the stress is transmitted from one particle to the other inside a chain through frictional or lubricated contacts. The factor 2 comes from the integral of the stress along a chain of length $\xi$. Note that the value $L$=1 gives a density profile with a cusp at the center (Phillips et al. 1992). In the frame of the W-C model Eq. (8) can be re-written as:

$$\frac{\Phi(r)}{\Phi_R} = \left(\frac{R}{r}\right)^L \sqrt{\frac{\eta(R)}{\eta(r)}} \tag{9}$$

With $\eta(r) = \alpha \left(1 - \frac{\Phi(r)}{\Phi_{J(r)}}\right)^{-2}$ \hfill (10)

In Eq. (10) the jamming volume fraction now will depend on r since in the capillary the stress:$\tau(r)=\tau_R$ (r/R) varies from zero at the center to $\tau_R$ on the wall so the jamming volume fraction will depend on r*=r/R :

$$\Phi_J(r^*) = (\Phi_{RLP} - \Phi_0)f(r^*\tau_R, \lambda, q) + \Phi_0 \tag{11}$$

Combining Eq. (9) and (10) gives the variation of the volume fraction:

$$\Phi(r^*, \tau_R) = \frac{\Phi_R}{r^{*l}\left(1 - \frac{\Phi_R}{\Phi_{JR}}\right) + \frac{\Phi_R}{\Phi_J(r^*)}} \tag{12}$$

Due to the divergence of the viscosity for $\Phi(r) = \Phi_J(r)$, the volume fraction in Eq. (12) remains bounded by $\Phi_J(r)$. The unknown in Eq. (12) is $\Phi_R$ the volume fraction at r=R. It is derived from the condition of recovering the average volume fraction:

$$\Phi = 2\int_0^1 \Phi(r^*, \tau_R)r^* dr^* \tag{13}$$

In Fig. (10) we have represented the volume fraction profile deduced from Eq. (12), (13) for four different stresses. These are equilibrium profiles when the migration is ended. For $\tau_R$=100 we are well below the critical stress and the maximum volume fraction at the center (r*=0) is equal to $\Phi_0$ and decreases almost linearly from the center to the wall. For $\tau_R$ =200, even if we are slightly above the transition, the profile is rather the same as for $\tau_R$ =100 except close to the wall where the volume fraction is slightly lower. The change is important at $\tau_R$ =275 Pa and still more important at $\tau_R$ =340, both values being above the jamming stress in the W-C model in the absence of migration. It is only at this last stress that the profile becomes blunted with, at the center of the capillary the maximum volume fraction Φ=Φ₀. This kind of profile was experimentally observed on suspensions of non-Brownian spheres with R/a ~40 at Φ=0.55 by magnetic resonance imaging (Oh et al. 2015).

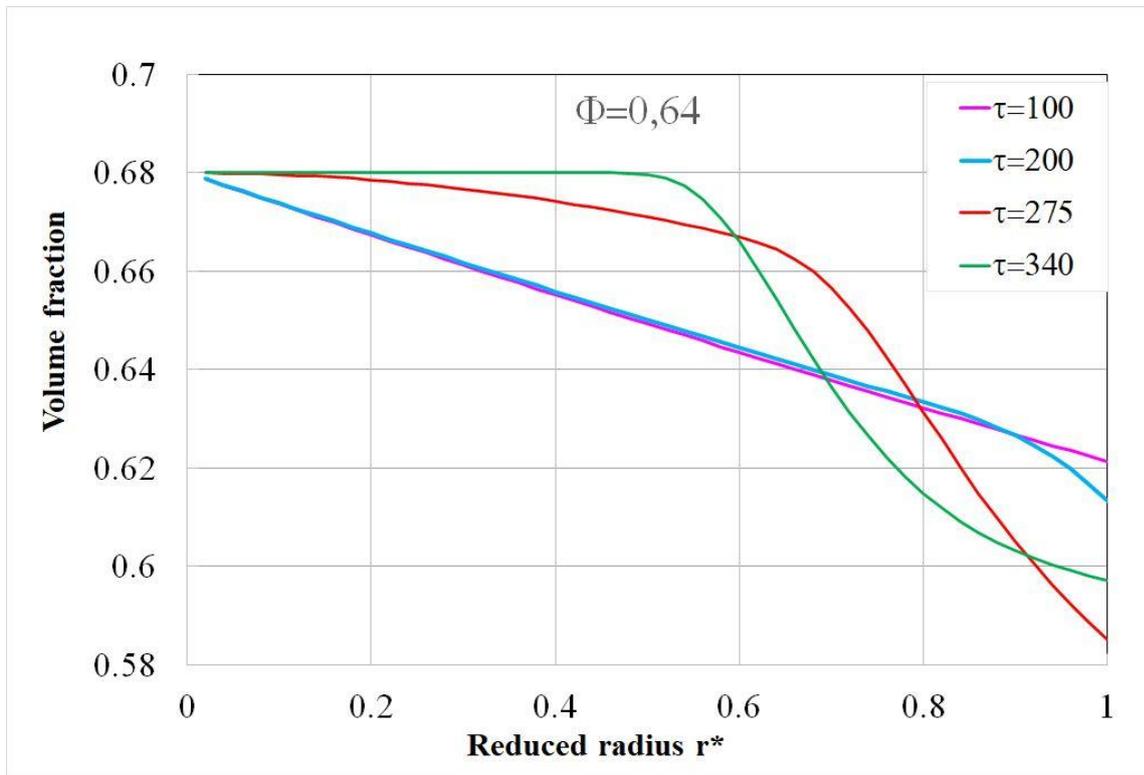

**Fig. (10))** Predicted volume fraction profiles Φ(r*) for four different wall stresses: τ=100Pa, τ=200Pa, τ=275Pa, τ=340Pa at an average volume fraction of Φ=0.64

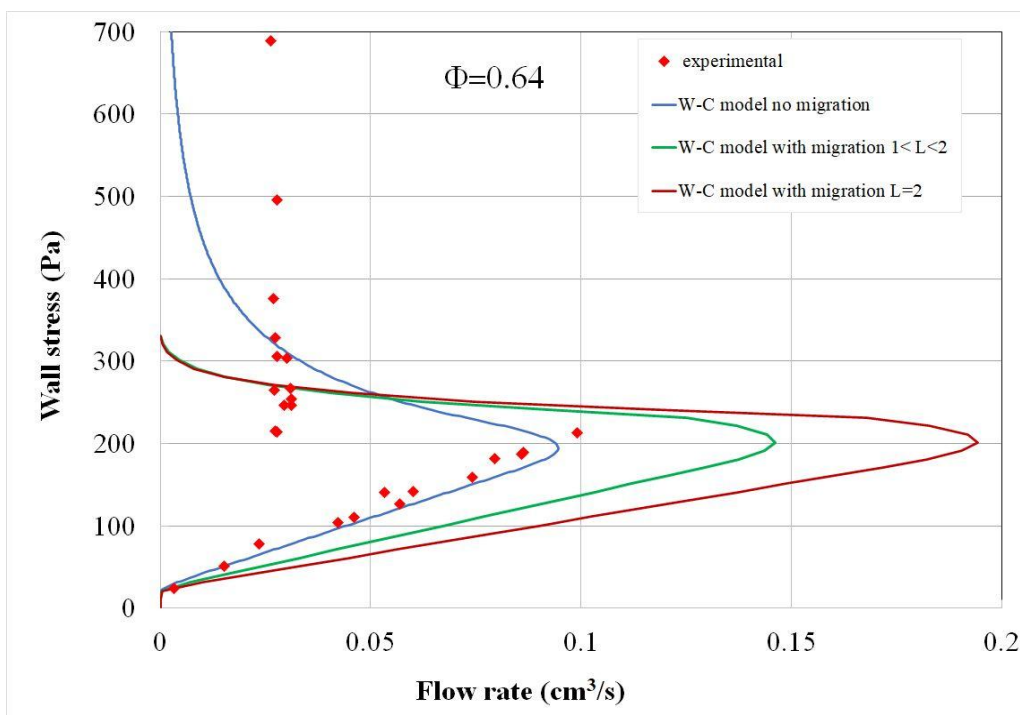

**Fig. (11)** Prediction of the stress versus flow rate in the frame of the W-C model. In blue without migration; in green with an exponent 1<L<2 (cf. Eq. (8)) depending on the value of the stress; in red with L=2 whatever the value of the stress. The experimental points are the red dots at Φ=0.64

The wall stress versus volume flow rate for different profiles of volume fraction, including the case of a constant volume fraction already presented in Fig. (7) in the absence of migration, is represented in Fig. (11)). In the presence of migration we use, instead of Eq. (4), the following one:

$$\dot{\gamma}(r, \tau_R) = \frac{\tau(r) - \tau_y}{\eta_r(\Phi, \tau)} \ with \ \eta_r(\Phi, r) = \alpha(1 - \frac{\Phi(r)}{\Phi_j(r)})^{-2}$$

where r= r*$\tau_R$ and Φ(r ) the solution of Eqs.(12),(13).

When the stress has exceeded the transition value, $\tau_c$, the jamming fraction $\Phi_J$ decreases with the stress until it reaches everywhere the local volume fraction Φ(r) making the flow to stop due to the divergence of the viscosity. We have taken in Eq. (8) a value of the exponent $L$=1+$f_c$($r^*$.$\tau_R$,λ,q)) which changes from 1 to 2 with the value of the stress to consider the progressive increase of the correlation length,$\xi$, with the stress. This correlation length is related to the presence of clusters of particles which transport the particles during their rotation on a length equivalent to the diameter of the cluster. Since, even in the absence of friction, the hydroclusters can also play this role, we can use $L$=2 also below the transition. In this case, we obtain the red curve which is still more different from the experimental results. It appears that, whatever the migration model, the increase in volume fraction at the center of the capillary will result in an important increase of the volume flow rate which is not compatible with the experimental data especially concerning the maximum volume flow rate. From this comparison between the models without and with migration and the experimental values we can conclude that we do not have migration.

References

Abbott JR, Tetlow N, Graham AL, et al (1991) Experimental observations of particle migration in concentrated suspensions: Couette flow. Journal of rheology 35:773–795

Baumgarten AS, Kamrin K (2019) A general constitutive model for dense, fine-particle suspensions validated in many geometries. Proceedings of the National Academy of Sciences 116:20828–20836

Bender J, Wagner NJ (1996) Reversible shear thickening in monodisperse and bidisperse colloidal dispersions. Journal of Rheology 40:899–916. https://doi.org/10.1122/1.550767

Bi D, Zhang J, Chakraborty B, Behringer RP (2011) Jamming by shear. Nature 480:355–358. https://doi.org/10.1038/nature10667

Bossis G, Brady JF (1989) The rheology of Brownian suspensions. The Journal of chemical physics 91:1866–1874

Bossis G, Boustingorry P, Grasselli Y, et al (2017) Discontinuous shear thickening in the presence of polymers adsorbed on the surface of calcium carbonate particles. Rheologica Acta 56:415–430

Bossis G, Volkova O, Grasselli Y, Gueye O (2019) Discontinuous shear thickening in concentrated suspensions. Philosophical Transactions of the Royal Society A 377:20180211

Bossis G, Grasselli Y, Ciffreo A, Volkova O (2021) Tunable discontinuous shear thickening in capillary flow of MR suspensions. Journal of Intelligent Material Systems and Structures 32:1349-1357

Brown E, Jaeger HM (2014) Shear thickening in concentrated suspensions: phenomenology, mechanisms and relations to jamming. Reports on Progress in Physics 77:046602

Chow AW, Sinton SW, Iwamiya JH, Stephens TS (1994) Shear-induced particle migration in Couette and parallel-plate viscometers: NMR imaging and stress measurements. Physics of Fluids 6:2561–2576

Clavaud C, Bérut A, Metzger B, Forterre Y (2017) Revealing the frictional transition in shear-thickening suspensions. Proceedings of the National Academy of Sciences 114:5147–5152

Comtet J, Chatté G, Niguès A, et al (2017) Pairwise frictional profile between particles determines discontinuous shear thickening transition in non-colloidal suspensions. Nature communications 8:1–7 d'Haene P, Mewis J, Fuller GG (1993) Scattering dichroism measurements of flow-induced structure of a shear thickening suspension. Journal of colloid and interface science 156:350–358

Da Cunha FR, Hinch EJ (1996) Shear-induced dispersion in a dilute suspension of rough spheres. Journal of fluid mechanics 309:211–223

Denn MM, Morris JF, Bonn D (2018) Shear thickening in concentrated suspensions of smooth spheres in Newtonian suspending fluids. Soft matter 14:170–184

Egres RG, Wagner NJ (2005) The rheology and microstructure of acicular precipitated calcium carbonate colloidal suspensions through the shear thickening transition. Journal of Rheology 49:719–746. https://doi.org/10.1122/1.1895800

Fagan ME, Zukoski CF (1997) The rheology of charge stabilized silica suspensions. Journal of Rheology 41:373–397. https://doi.org/10.1122/1.550876

Fall A, Huang N, Bertrand F, et al (2008) Shear Thickening of Cornstarch Suspensions as a Reentrant Jamming Transition. Phys Rev Lett 100:018301. https://doi.org/10.1103/PhysRevLett.100.018301

Fall A, Lemaitre A, Bertrand F, et al (2010) Shear thickening and migration in granular suspensions. Physical review letters 105:268303

Fall A, Bertrand F, Hautemayou D, et al (2015) Macroscopic discontinuous shear thickening versus local shear jamming in cornstarch. Physical review letters 114:098301

Franks GV, Zhou Z, Duin NJ, Boger DV (2000) Effect of interparticle forces on shear thickening of oxide suspensions. Journal of Rheology 44:759–779. https://doi.org/10.1122/1.551111

Frith WJ, d'Haene P, Buscall R, Mewis J (1996) Shear thickening in model suspensions of sterically stabilized particles. Journal of Rheology 40:531–548. https://doi.org/10.1122/1.550791

Gameiro M, Singh A, Kondic L, et al (2020) Interaction network analysis in shear thickening suspensions. Phys Rev Fluids 5:034307. https://doi.org/10.1103/PhysRevFluids.5.034307

Graham AL, Altobelli SA, Fukushima E, et al (1991) Note: NMR imaging of shear-induced diffusion and structure in concentrated suspensions undergoing Couette flow. Journal of Rheology 35:191–201

Guy BM, Ness C, Hermes M, et al (2020) Testing the Wyart–Cates model for non-Brownian shear thickening using bidisperse suspensions. Soft matter 16:229–237

Hampton RE, Mammoli AA, Graham AL, et al (1997) Migration of particles undergoing pressure-driven flow in a circular conduit. Journal of Rheology 41:621–640

Hermes M, Guy BM, Poon WC, et al (2016) Unsteady flow and particle migration in dense, non-Brownian suspensions. Journal of Rheology 60:905–916

Hoffman RL (1972) Discontinuous and Dilatant Viscosity Behavior in Concentrated Suspensions. I. Observation of a Flow Instability. Transactions of the Society of Rheology 16:155–173. https://doi.org/10.1122/1.549250

Hsu C-P, Ramakrishna SN, Zanini M, et al (2018) Roughness-dependent tribology effects on discontinuous shear thickening. Proceedings of the National Academy of Sciences 115:5117–5122

Isa L, Besseling R, Poon WC (2007) Shear zones and wall slip in the capillary flow of concentrated colloidal suspensions. Physical Review Letters 98:198305

Isa L, Besseling R, Morozov AN, Poon WC (2009) Velocity oscillations in microfluidic flows of concentrated colloidal suspensions. Physical review letters 102:058302

Jerkins M, Schröter M, Swinney HL, et al (2008) Onset of Mechanical Stability in Random Packings of Frictional Spheres. Phys Rev Lett 101:018301. https://doi.org/10.1103/PhysRevLett.101.018301

Johnson DH, Vahedifard F, Jelinek B, Peters JF (2017) Micromechanical modeling of discontinuous shear thickening in granular media-fluid suspension. Journal of Rheology 61:265–277

Koivisto J, Durian DJ (2017) Effect of interstitial fluid on the fraction of flow microstates that precede clogging in granular hoppers. Physical Review E 95:032904

Laun HM, Bung R, Schmidt F (1991) Rheology of extremely shear thickening polymer dispersionsa)(passively viscosity switching fluids). Journal of Rheology 35:999–1034

Laun HM, Bung R, Hess S, et al (1992) Rheological and small angle neutron scattering investigation of shear-induced particle structures of concentrated polymer dispersions submitted to plane Poiseuille and Couette flowa). Journal of Rheology 36:743–787. https://doi.org/10.1122/1.550314

Laun HM (1994) Normal stresses in extremely shear thickening polymer dispersions. Journal of non-newtonian fluid mechanics 54:87–108

Lee Y-F, Luo Y, Brown SC, Wagner NJ (2020) Experimental test of a frictional contact model for shear thickening in concentrated colloidal suspensions. Journal of Rheology 64:267–282

Leighton D, Acrivos A (1987) The shear-induced migration of particles in concentrated suspensions. Journal of Fluid Mechanics 181:415–439

Madraki Y, Oakley A, Nguyen Le A, et al (2020) Shear thickening in dense non-Brownian suspensions: Viscous to inertial transition. Journal of Rheology 64:227–238

Mari R, Seto R, Morris JF, Denn MM (2014) Shear thickening, frictionless and frictional rheologies in non-Brownian      suspensions. Journal of Rheology 58:1693–1724. https://doi.org/10.1122/1.4890747

Meunier A, Bossis G (2008) The influence of surface forces on shear-induced tracer diffusion in mono and bidisperse suspensions. The European Physical Journal E 25:187–199

Mills P, Snabre P (1995) Rheology and structure of concentrated suspensions of hard spheres. Shear induced particle migration. Journal de Physique II 5:1597–1608

More RV, Ardekani AM (2020) A constitutive model for sheared dense suspensions of rough particles. Journal of Rheology 64:1107–1120

Morris JF (2020) Shear thickening of concentrated suspensions: Recent developments and relation to other phenomena. Annual Review of Fluid Mechanics 52:121–144

Nakanishi H, Nagahiro S, Mitarai N (2012) Fluid dynamics of dilatant fluids. Physical Review E 85:011401

Neuville M, Bossis G, Persello J, et al (2012) Rheology of a gypsum suspension in the presence of different superplasticizers. Journal of Rheology 56:435–451. https://doi.org/10.1122/1.3693272

Oh S, Song Y, Garagash DI, et al (2015) Pressure-driven suspension flow near jamming. Physical review letters 114:088301

Phillips RJ, Armstrong RC, Brown RA, et al (1992) A constitutive equation for concentrated suspensions that accounts for shear-induced particle migration. Physics of Fluids A: Fluid Dynamics 4:30–40

Quemada D, Berli C (2002) Energy of interaction in colloids and its implications in rheological modeling. Advances in Colloid and Interface Science 98:51–85. https://doi.org/10.1016/S0001-8686(01)00093-8

Richards JA, Royer JR, Liebchen B, et al (2019) Competing Timescales Lead to Oscillations in Shear-Thickening Suspensions. Phys Rev Lett 123:038004. https://doi.org/10.1103/PhysRevLett.123.038004

Richards JA, Guy BM, Blanco E, et al (2020) The role of friction in the yielding of adhesive non-Brownian suspensions. Journal of Rheology 64:405–412. https://doi.org/10.1122/1.5132395

Seto R, Mari R, Morris JF, Denn MM (2013) Discontinuous shear thickening of frictional hard-sphere suspensions. Physical review letters 111:218301

Singh A, Mari R, Denn MM, Morris JF (2018) A constitutive model for simple shear of dense frictional suspensions. Journal of Rheology 62:457–468. https://doi.org/10.1122/1.4999237

Singh A, Pednekar S, Chun J, et al (2019) From yielding to shear jamming in a cohesive frictional suspension. Physical review letters 122:098004

Wyart M, Cates ME (2014) Discontinuous shear thickening without inertia in dense non-Brownian suspensions. Physical review letters 112:098302

Zarraga IE, Leighton Jr DT (2001) Shear-induced diffusivity in a dilute bidisperse suspension of hard spheres. Journal of colloid and interface science 243:503–514